\documentclass[sigconf,10pt]{acmart}
\usepackage{graphicx}
\usepackage{ulem}
\usepackage{cancel}
\usepackage{amsthm}
\usepackage{amsmath}
\usepackage{amsfonts}
\usepackage{amsmath}
\usepackage{graphicx}
\usepackage{mathtools}
\usepackage{multirow}
\usepackage{epstopdf}
\usepackage{subfigure}
\usepackage{array}
\usepackage{csquotes}
\usepackage{lipsum}
\usepackage{algorithmic}
\usepackage{caption}
\graphicspath{{Figures/}}
\usepackage{mathrsfs} 
\usepackage{listings}
\usepackage{amsmath,relsize,mathtools}
\usepackage{lipsum}

\usepackage{xcolor}
\usepackage[linesnumbered,ruled,lined,boxed]{algorithm2e}

\SetKwInput{KwInput}{Input}                
\SetKwInput{KwOutput}{Output}              

\newcolumntype{C}[1]{>{\centering\let\newline\\\arraybackslash\hspace{0pt}}m{#1}}

\makeatother

\begin{document}

\acmYear{2023}\copyrightyear{2023}
\setcopyright{acmlicensed}
\acmConference[ACM WiNTECH' 23]{The 17th ACM Workshop on Wireless Network Testbeds, Experimental evaluation \& Characterization 2023}{October 2--6, 2023}{Madrid, Spain}
\acmBooktitle{The 17th ACM Workshop on Wireless Network Testbeds, Experimental evaluation \& Characterization 2023 (ACM WiNTECH' 23), October 2--6, 2023, Madrid, Spain}
\acmPrice{15.00}
\acmDOI{10.1145/3570361.3613199}
\acmISBN{978-1-4503-9990-6/23/10}


\title[An Extensive Measurement Campaign of A Dense Indoor Wi-Fi 6E Network]{Evaluating The Interference Potential in 6 GHz: An Extensive Measurement Campaign of A Dense Indoor Wi-Fi 6E Network}

\author{Seda Dogan-Tusha}
\affiliation{%
  \institution{University of Notre Dame}
  \country{}
}
\email{stusha@nd.edu}

\author{Muhammad I. Rochman}
\affiliation{%
  \institution{University of Chicago}
  \country{}
}
\email{muhiqbalcr@uchicago.edu}

\author{Armed Tusha}
\affiliation{%
  \institution{University of Notre Dame}
  \country{}
}
\email{atusha@nd.edu}

\author{Hossein Nasiri}
\affiliation{%
  \institution{University of Notre Dame}
  \country{}
}
\email{hnasiri2@nd.edu}

\author{James Helzerman}
\affiliation{%
  \institution{University of Michigan}
  \country{}
}
\email{jarhelz@umich.edu}

\author{Monisha Ghosh}
\affiliation{%
  \institution{University of Notre Dame}
  \country{}
}
\email{mghosh3@nd.edu}

\renewcommand{\shortauthors}{S. Tusha, \textit{et al.}}

\begin{abstract}
 
The Federal Communications Commission (FCC) has allocated the 6 GHz band (5.925 - 7.125 GHz) for unlicensed, shared use in the US. Incumbents in the band are protected via Low Power Indoor (LPI) rules that do not require the use of 
an Automatic Frequency Control (AFC) mechanism and Standard Power (SP) rules which do. 
As the deployment of Wi-Fi 6E APs implementing LPI rules have been increasing, 
there is limited research examining the real-world interference potential of dense LPI deployments to fixed links, which remains a concern for incumbents. We have conducted a first-of-its-kind extensive measurement campaign of a dense indoor Wi-Fi 6E network at the University of Michigan, which includes walking, driving, and drone measurements to assess outdoor beacon Received Signal Strength Indicator (RSSI), building entry loss (BEL), channel utilization, and appropriate enabling signal level for a proposed client-to-client (C2C) mode in 6 GHz. 
Our detailed measurements under various conditions show median outdoor RSSI between -75 dBm and -85 dBm, BEL between 12 dB and 16 dB through double-pane low-emission windows, and only $5\%$ of indoor Basic Service Set Identifiers (BSSIDs) observed outdoors. Our overall conclusion is that the probability of interference to incumbent fixed links is low, but more research is required to determine the appropriate signal level for the C2C enabling signal.
\end{abstract}

\maketitle

\section{Introduction}\label{introduction}

\subsection{Unlicensed use of 6 GHz in the US}

The increasing bandwidth demands of new wireless applications and use cases prompted the U.S. Federal Communications Commission (FCC), in 2020, to allocate the 6 GHz band from 5.925 GHz to 7.125 GHz for unlicensed use on a shared basis with existing incumbents, primarily fixed microwave links, cable television relay services (CTRS), satellite and mobile Broadcast Auxiliary Services (BAS) \cite{FCC1}. While some countries have only allocated the lower 500 MHz on an unlicensed basis reserving the upper portion for possible future auctions and licensing \cite{WiFi2}, the large number of incumbents in the band ($>$ 48,000) made the prospect of relocating incumbents prior to licensing a major challenge for the U.S. Hence, the most expedient way of making this band available for commercial applications was to develop rules for unlicensed devices to use this band while sharing with incumbents.
Since the majority of wireless traffic, approximately 60\%, is handled by Wi-Fi \cite{9762843}, 
allocating this band for unlicensed use also relieves the growing congestion in the existing 2.4 GHz and 5 GHz unlicensed bands. 

\begin{figure}
    \centering
    \includegraphics[width=\linewidth]{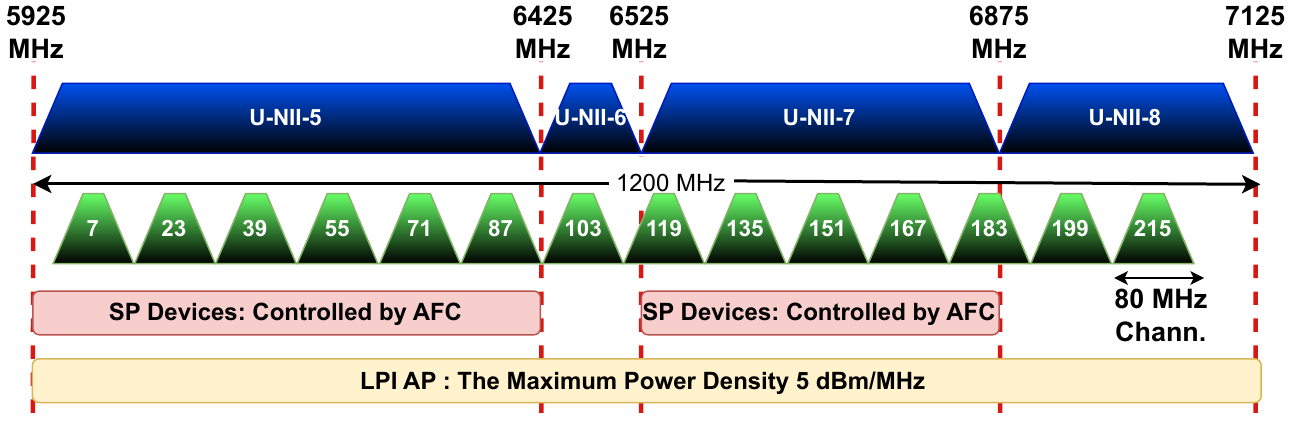}
    \vspace{-2em}
    \caption{Channelization of the 6 GHz band.}
    \label{Fig:6ghz_freq_chart}
    \vspace{-1.5em}
\end{figure}

The 6 GHz band encompasses four U-NII (Unlicensed National Information and Infrastructure) bands: U-NII-5 to U-NII-8, as listed with its incumbents in Table~\ref{tab:UNII_6e}. 
Incumbents are protected via two sets of rules that unlicensed devices must follow: low power indoor (LPI) and standard power (SP).
LPI operation is permitted across the entire 6 GHz band without the need for an Automatic Frequency Control (AFC) system, but access points (APs) must be installed indoors. SP APs can be installed anywhere, but are limited to U-NII-5 and U-NII-7 and require an AFC to avoid interference with incumbents. 
Thus, Wi-Fi 6E devices can utilize 14 additional 80 MHz channels and 7 additional 160 MHz channels over the total 1200 MHz span 
as shown in Fig. \ref{Fig:6ghz_freq_chart}. This paper focuses on LPI deployments since the AFC is still under development and SP APs have not yet been deployed.

Unlike the 5 GHz band regulations, Wi-Fi 6E APs operating under LPI rules in the 6 GHz band must adhere to a maximum power spectral density (PSD) of 5 dBm/MHz, regardless of the channel bandwidth \cite{vanlin1}. 
This corresponds to maximum transmit (Tx) powers shown in Table \ref{tab:regulation_6e}. While APs are limited to indoor use, client devices (STAs) can be anywhere, including outdoors, and are therefore required to transmit 6 dB less power than the AP.

In the Further Notice of Proposed Rulemaking (FNPRM) \cite{FCC1} and the Public Notice \cite{PN}, the FCC is considering enhancing the 6 GHz rules in the future by (i) raising the PSD limit to 8 dBm/MHz, (ii) adding a Very Low Power (VLP) option with a PSD of -8 dBm/MHz and maximum transmit power of 14 dBm that can be used anywhere without requiring AFC access, and (iii) implementing a client-to-client (C2C) mode that enables direct connections between client devices via an enabling signal from a LPI AP. These options require further research to ensure that incumbents will continue to be protected: the results of this paper will inform this process.

\vspace{-.2em}

\subsection{Related Work}

While various coexistence scenarios in 6 GHz have been studied in the academic literature, there is a lack of research on analyzing interference to incumbents. In \cite{9488780}, the use of multi-user orthogonal frequency division multiple access (MU-OFDMA) for uplink Wi-Fi 6E is proposed for coexistence with 3GPP-based unlicensed technologies, i.e., 5G NR-U. 
Authors in \cite{9826115}  consider the impact of Wi-Fi 6E on ultra-wideband (UWB) communications and ranging.
In \cite{9722698}, the authors study the adjacent channel interference between Wi-Fi 6E in 6 GHz and C-V2X (Cellular Vehicle-to-Everything) in the adjacent 5.9 GHz.

\begin{table}
	\caption{{Unlicensed Operation over 6 GHz.}}
    \vspace{-1em}
	\centering
	 \footnotesize
\renewcommand{\arraystretch}{1}
	\begin{tabular}{|C{1 cm}|C{2 cm}|C{1 cm}| C{ 1 cm}| C{ 1.5cm}|} 
 \hline
	 \textbf{Band} & \textbf{Incumbents}    &  \textbf{Use Cases} & \textbf{ Chann. No.} &
	 \textbf{Freq. (MHz)}	 \\ 	\hline \hline
U-NII-5	 & Fixed, Satellite Uplink  & LPI, SP & 1-97 & 5925-6425 \\
		\hline 
	U-NII-6&  Satellite uplink, BAS, CTRS & LPI  & 101-117& 6425-6525 \\
		\hline
U-NII-7   &   Fixed, Satellite uplink/downlink & LPI, SP & 121-185 & 6525-6875 \\
            \hline
 U-NII-8      &  Fixed, Satellite, BAS  & LPI  & 189-233 & 6875-7125 \\
            \hline
	\end{tabular}
	\label{tab:UNII_6e} 
    \vspace{-1em}
\end{table}

\begin{table}
	\caption{{Max. Tx Power for 6 GHz LPI.}}
    \vspace{-1em}
	\centering
	\footnotesize
	\renewcommand{\arraystretch}{1}
	\begin{tabular}{|C{1cm}|*{5}{C{1.05 cm}|}} 
 \hline
	\textbf{Device} &  \multicolumn{5}{c|}{\textbf{ Maximum TX Power}}

		 \\ 	\cline{2-6}
	 \textbf{Type} & $20$ MHz & $40$ MHz & $80$ MHz & $160$ MHz & $320$ MHz\\
		 \hline \hline
	STA &  12 dBm & 15 dBm  & 18 dBm& 21 dBm & 24 dBm\\
		\hline
     AP &   18 dBm  & 21 dBm  & 24 dBm & 27 dBm  & 30 dBm\\
            \hline
	\end{tabular}
	\label{tab:regulation_6e} 
    \vspace{-1.5em}
\end{table}

\begin{figure*}
     \centering
     \begin{subfigure}[Wi-Fi deployment at UMich. Green Pins: Buildings with Wi-Fi 6E LPI APs.]{  \centering
    \includegraphics[width=0.28\textwidth]{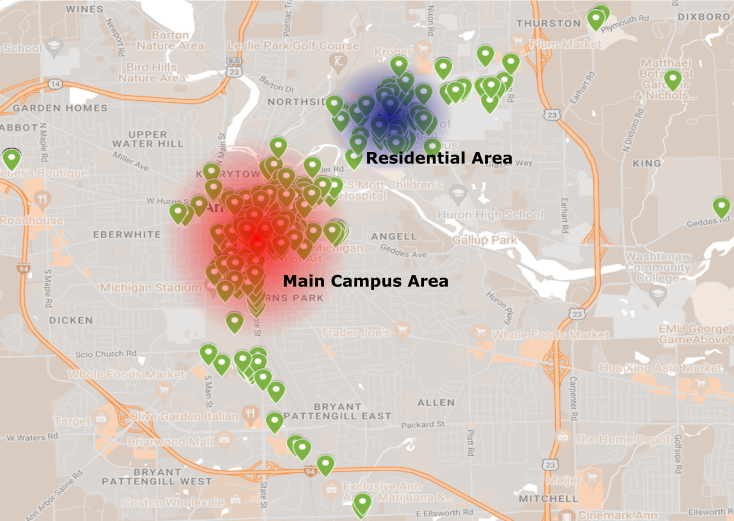}
         \label{Fig:wifi_deployment}}
    \end{subfigure}
     \begin{subfigure}[Driving and walking meas. in the MCA. Red Pins: Driving, Blue Pins: Walking.]{
         \centering
\includegraphics[width=0.28\textwidth]{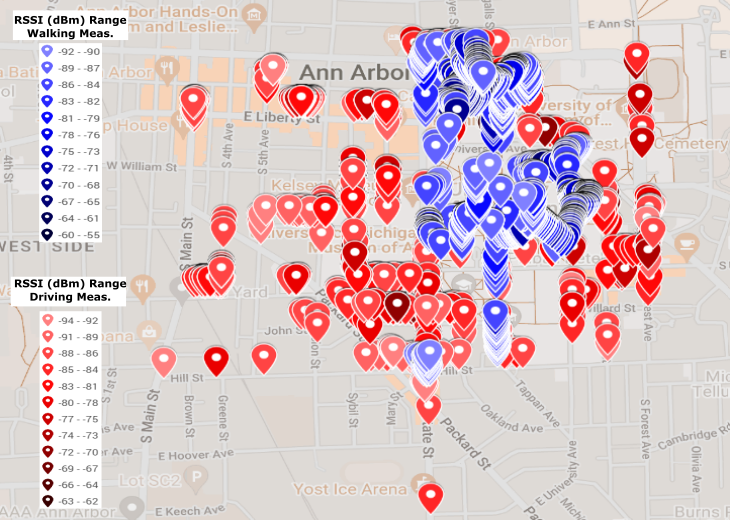}
        \label{Fig:reg_campus1}}
     \end{subfigure} 
     \begin{subfigure}[6 GHz fixed links. Orange: drone meas. locations, Blue: Tx locations, Red: Rx location.]{
         \centering
\includegraphics[width=0.28\textwidth]{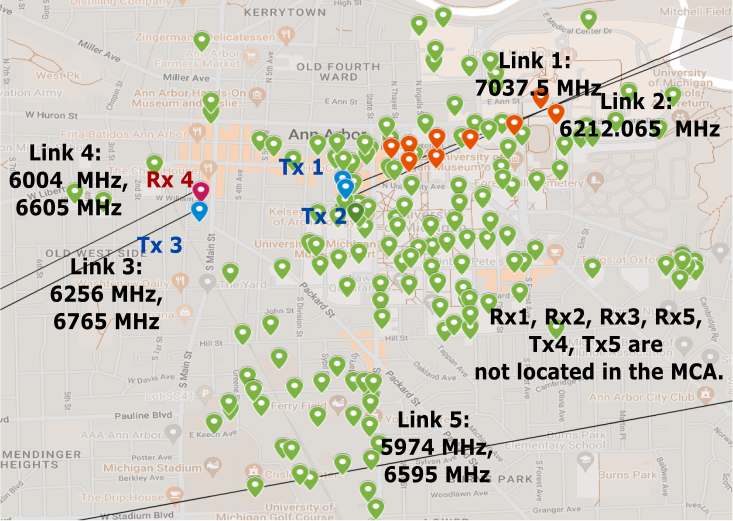}
        \label{Fig:reg_campus2}}
     \end{subfigure}
    \vspace{-1em}
    \caption{The main campus area (MCA) and the residential area (RA) in UMich.}
    \label{Fig:meas_locs}
    \vspace{-1em}
\end{figure*}

Studies of coexistence with incumbents are mainly conducted by various industry stakeholders, particularly fixed links operators and unlicensed proponents. 
In \cite{epri1, epri} the additive impact of unlicensed LPI operations is assessed by analyzing Wi-Fi 6E APs operating co-channel within the main beam of an operational 
system in the FirstEnergy network. In \cite{pge}, a predicted potential interference analysis over 5 years has been presented for Pacific Gas \& Electric’s deployments. In \cite{evergy, evergy2}, the fade margin degradation of a 6 GHz link in the presence of a Wi-Fi 6E AP in the path is demonstrated. However, the experimental set-ups in studies by the incumbents are quite contrived, e.g., APs intentionally placed near windows on the same channel as an incumbent and within the main beam, which are not reflected in real-world deployments. Hence our goal is to understand the statistics of interference based on a dense real-world deployment, instead of worst-case scenarios.

\subsection{Motivation \& Main Contributions}

Given the above discussion, the aim of this study is to evaluate, in an unbiased manner, the potential for interference to outdoor fixed links from a real-world, densely deployed 6 GHz LPI network. The main contributions of this paper are:

\noindent $\bullet$
A first-of-its-kind, extensive measurement campaign undertaken on the main campus of the University of Michigan (UMich) in Ann Arbor which has more than 16,000 Wi-Fi 6E LPI APs deployed across 225 buildings. This is the largest such deployment in the world today.

\noindent $\bullet$
Generating heat-maps at ground level of outdoor Received Signal Strength Indicator (RSSI) measured on the 20 MHz beacon frames transmitted by LPI APs, using measurements obtained by walking and driving on the main campus area (MCA) and the nearby residential area (RA). 

\noindent $\bullet$
Drone measurements around buildings near the path of 6 GHz fixed links to assess outdoor RSSI levels at higher altitudes where these links are deployed.

\noindent $\bullet$
We demonstrate median outdoor RSSI levels of -82 dBm while driving and -77 dBm while walking on campus. We show that the number of APs within a building, the positioning of the APs in relation to nearby windows, construction type and window materials, all are crucial in determining the outdoor RSSI levels. 

\noindent $\bullet$
Despite the significant number of deployed indoor APs, each with an average of two Basic Service Set Identifiers (BSSIDs) per AP, only 5\% of these BSSIDs are observed outdoors thus indicating that the potential for interference is limited to a smaller number than the deployed number.

\noindent $\bullet$
Measurements of building loss in two buildings on campus demonstrate building entry loss (BEL) of 12 - 16 dB through double-pane low-E windows. 

\section{Tools and Methodology}

 Fig. \ref{Fig:wifi_deployment} shows the Wi-Fi 6E deployments in the MCA and the RA of UMich. The campus is located in Ann Arbor with a high density of pedestrian and vehicular traffic, serving as an ideal location to assess potential interference from densely deployed indoor 6 GHz networks. The majority of the buildings in the MCA have double-pane low-E windows. Only 227 APs are deployed in the RA, which is a less dense deployment compared to the MCA which has a few thousand deployed APs.
\subsection{Measurement Tools}
Client devices were used to capture signal information in various environments, using two tools, SigCap and Wireshark, on smartphones and laptops respectively, to extract various signal parameters as shown in Table \ref{tab:features_devices}.

{\it \bf SigCap} is a custom Android app that passively collects time and geo-stamped wireless signal parameters (cellular and Wi-Fi) through APIs without root access \cite{9261954}. Wi-Fi parameters, such as RSSI, channel, BSSID, etc. are collected from the beacon frames every 5 seconds. Optional beacon elements with information on Tx signal power, number of stations connected to each BSSID and channel utilization (percentage of time that the AP senses the channel to be busy) are also collected:  fortunately, all the Wi-Fi 6E APs deployed in UMich broadcast these optional elements, thus facilitating our analysis. {\it \bf Wireshark} is an open source tool that we used for capturing both beacon and data frames using a Lenovo ThinkPad P16 Gen1 with the Intel(R) Wi-Fi AX211 Wi-Fi adapter.

We intentionally did not use spectrum analyzers for this work since the bursty nature of Wi-Fi traffic and the low outdoor RSSI levels are better captured using the above tools. Using smartphones enables mobile data collection which is difficult to do even with a handheld spectrum analyzer.

\subsection{Methodology}
The measurements were conducted in two campaigns, as described below.

\subsubsection{Measurement Campaign 1 (MC1):}
MC1 took place on January 7-9, 2023, during which measurements were conducted while driving and in a fixed location on campus. 

\begin{table} [h]
	\caption{Measurement tools and devices.}
    \vspace{-1em}
	\centering
	\footnotesize
        \renewcommand{\arraystretch}{1}
	\begin{tabular}{|C{1.5cm}|C{2.9cm}|C{3cm}|} 
 \hline
 \textbf{Tool} & \textbf{Wi-Fi Parameters}  & \textbf{Devices}\\
  \hline \hline  \textbf{SigCap} &  Time-stamp, location,
  frequency, RSSI, BSSID, SSID, $\#$STA, Channel Utilization & 1 $\times$ Google Pixel 6,\newline 1 $\times$ Samsung S21 Ultra,\newline 3 $\times$ Samsung S22+\\
  \hline
  \textbf{Wireshark} & Source/Destination, SSID, BSSID, Frequency, RSSI, Tx Power, beacon and data packets & Laptop: ThinkPad P16 Gen 1, Wi-Fi Card: Intel(R) Wi-Fi 6E AX211 160 MHz, OS: Ubuntu 22.04 LTS  \\
  \hline
	\end{tabular} 
	\label{tab:features_devices} 
    \vspace{-1em}
\end{table}
\begin{figure}
     \centering
     \begin{subfigure}[Indoor location and Wi-Fi 6E AP.]{  \centering
    \includegraphics[width=0.46\linewidth]{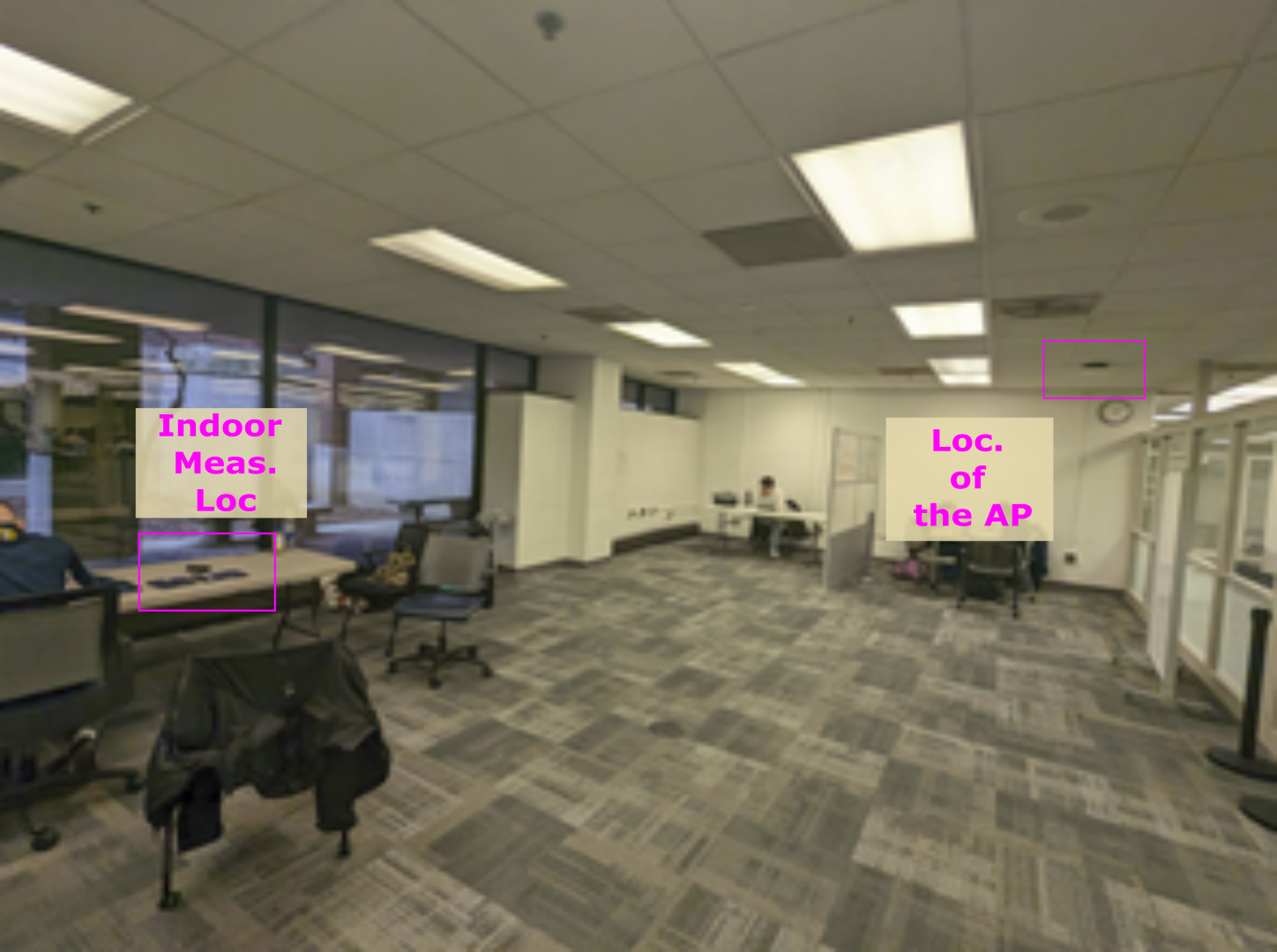}
         \label{Fig:ShapiroAP}}
    \end{subfigure}
    \hfill
     \begin{subfigure}[Outdoor locations.]{
         \centering
         \includegraphics[width=0.46\linewidth]{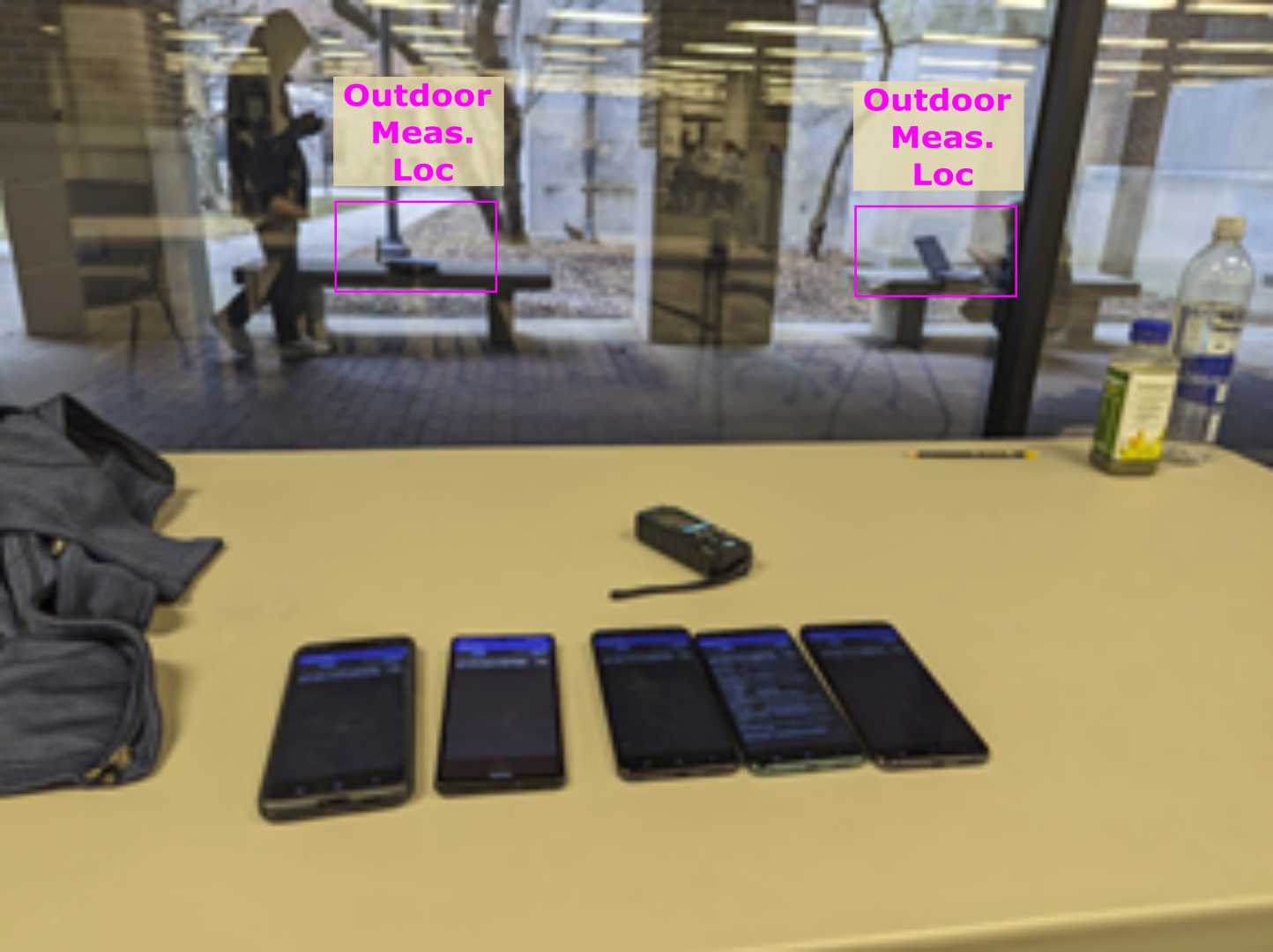}
        \label{Fig:ShapiroME}}
     \end{subfigure} 
    \vspace{-1em}
    \caption{Measurement and AP locations for FL1.}
    \label{Fig:meas_locs_jan}
    \vspace{-1em}
\end{figure}

\textit{Driving Measurements} were conducted in the MCA as shown in Fig.~\ref{Fig:reg_campus1} between 9:50 pm and 00:50 am, at a speed of 20 miles per hour. Data was collected with SigCap running on the five smartphones listed in Table \ref{tab:features_devices}. Due to the cold weather, walking measurements were not conducted in MC1.

\textit{Fixed Location 1 (FL1)} measurements were taken inside and outside a building with an open indoor area with high occupancy. Fig. \ref{Fig:ShapiroAP} shows the position of the Wi-Fi 6E LPI AP in the space. The AP is positioned 6 meters away from double pane low-E windows. The indoor measurements were taken by placing the phones near the window while the outdoor measurement location is 1.5 meters from the window. 
Wireshark and Sigcap were both used for measurements, as shown in Fig.~\ref{Fig:ShapiroME}. The AP transmit power was 15 dBm over a 160 MHz channel bandwidth, which is considerably lower than the regulatory limits specified in Table \ref{tab:regulation_6e}. This reduction in transmit power is due to the dense deployment of LPI APs, since many users need to be supported in this area.

\begin{figure}
     \centering
    { \includegraphics[width=0.92\linewidth]{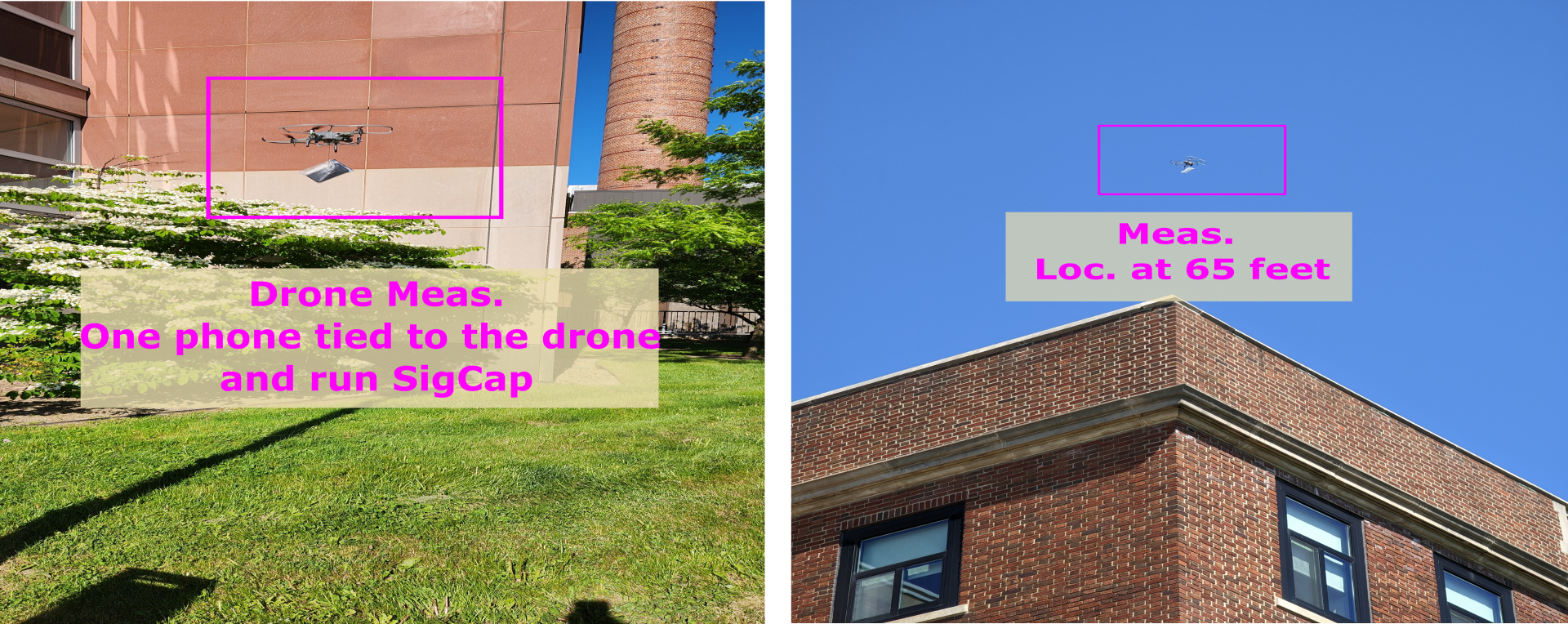}} 
    \vspace{-1em}
    \caption{Drone measurement scenario.}
    \label{Fig:meas_locs_drone}
    \vspace{-1em}
\end{figure}

\subsubsection{Measurement Campaign 2 (MC2):} MC2 was conducted on May 24-27, 2023 with drone, driving, walking, and fixed location measurements. The deployment had been changed from mostly 160 MHz channels observed in MC1 to mostly 80 MHz channels during MC2. This was done by UMich Information and Technology Services (ITS) to serve more users with a higher quality of service. However, since our analysis depends on measurements of the 20 MHz beacon channel RSSI, this change did not affect our results or comparisons.

\begin{table}
	\caption{Building information for drone measurements.}
    \vspace{-1em}
	\centering
	\footnotesize
        \renewcommand{\arraystretch}{1}
	\begin{tabular}{|C{3 cm}|C{2cm}|C{2cm}|} 
 \hline
 \textbf{Building Name} & \textbf{Height (ft)}  & \textbf{No. of AP/BSSIDs}\\
  \hline
  \hline
  Building 1 (BLD1) & 58 & 43/86 \\
  \hline
  Building 2 (BLD2) & 40 & 184/368 \\
  \hline
  Building 3 (BLD3) & 45 & 44/88 \\
  \hline 
  Building 4 (BLD4) & 47-65 wrt. upper and lower levels& 400/800 \\
  \hline
  Building 5 (BLD5) & 58&39/78\\
  \hline
  Building 6 (BLD6) & 65-85 wrt. upper and lower levels& 46/92\\
  \hline 
   Building 7 (BLD7) & 75 & 40/80\\
  \hline
  Building 8 (BLD8) & 70& 40/80\\
  \hline 
  Building 9 (BLD9) & 70& 40/80\\
 \hline
	\end{tabular}
	\label{tab:dronebuildinginfo} 
    \vspace{-1.5em}
\end{table}

\begin{figure}
     \centering
     \begin{subfigure}[Indoor locations and Wi-Fi 6E AP]{  \centering
    \includegraphics[width=0.46\linewidth]{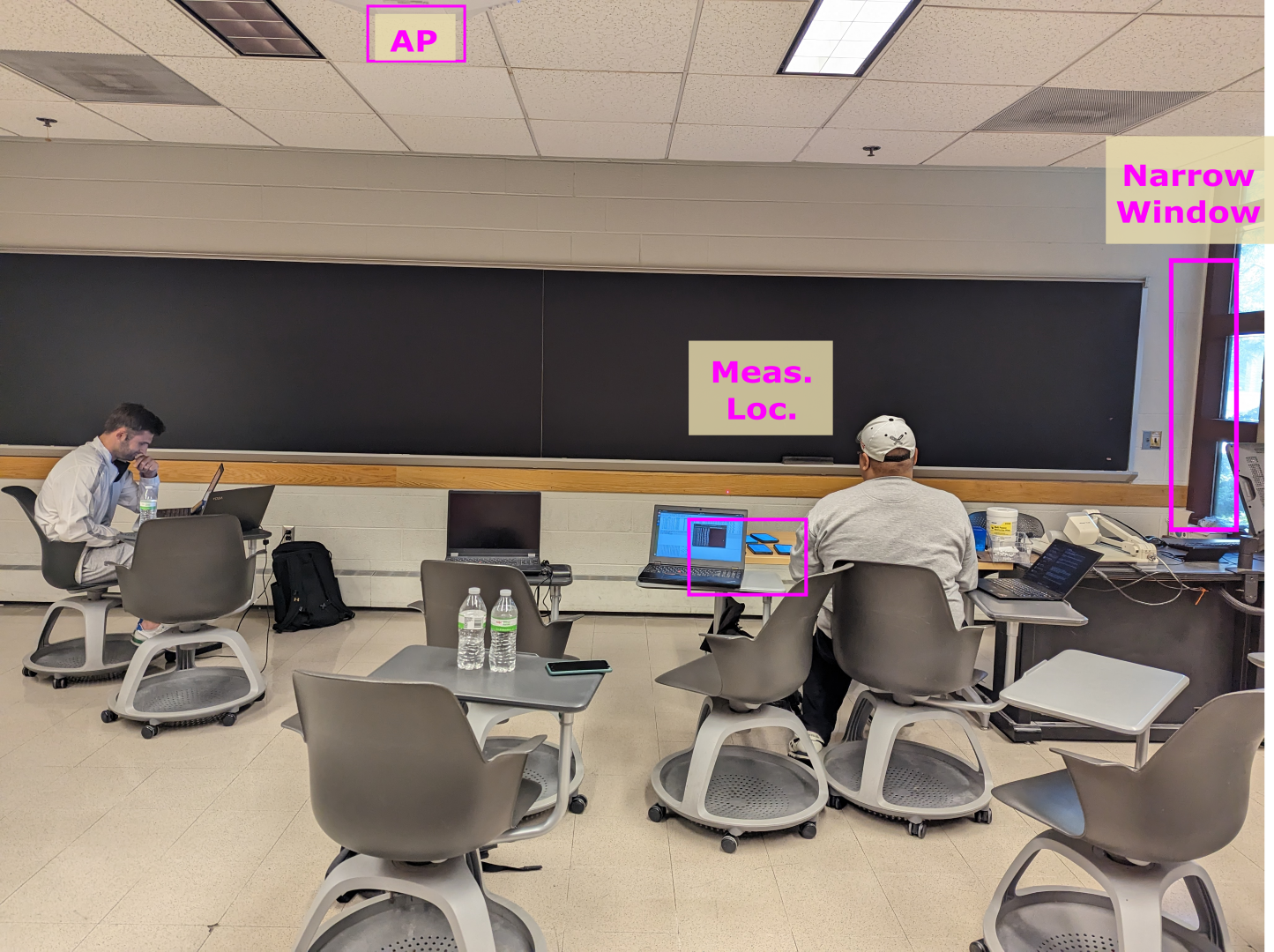}
         \label{Fig:chemistryin}}
    \end{subfigure} \hfill
     \begin{subfigure}[Outdoor locations]{
         \centering
         \includegraphics[width=0.46\linewidth]{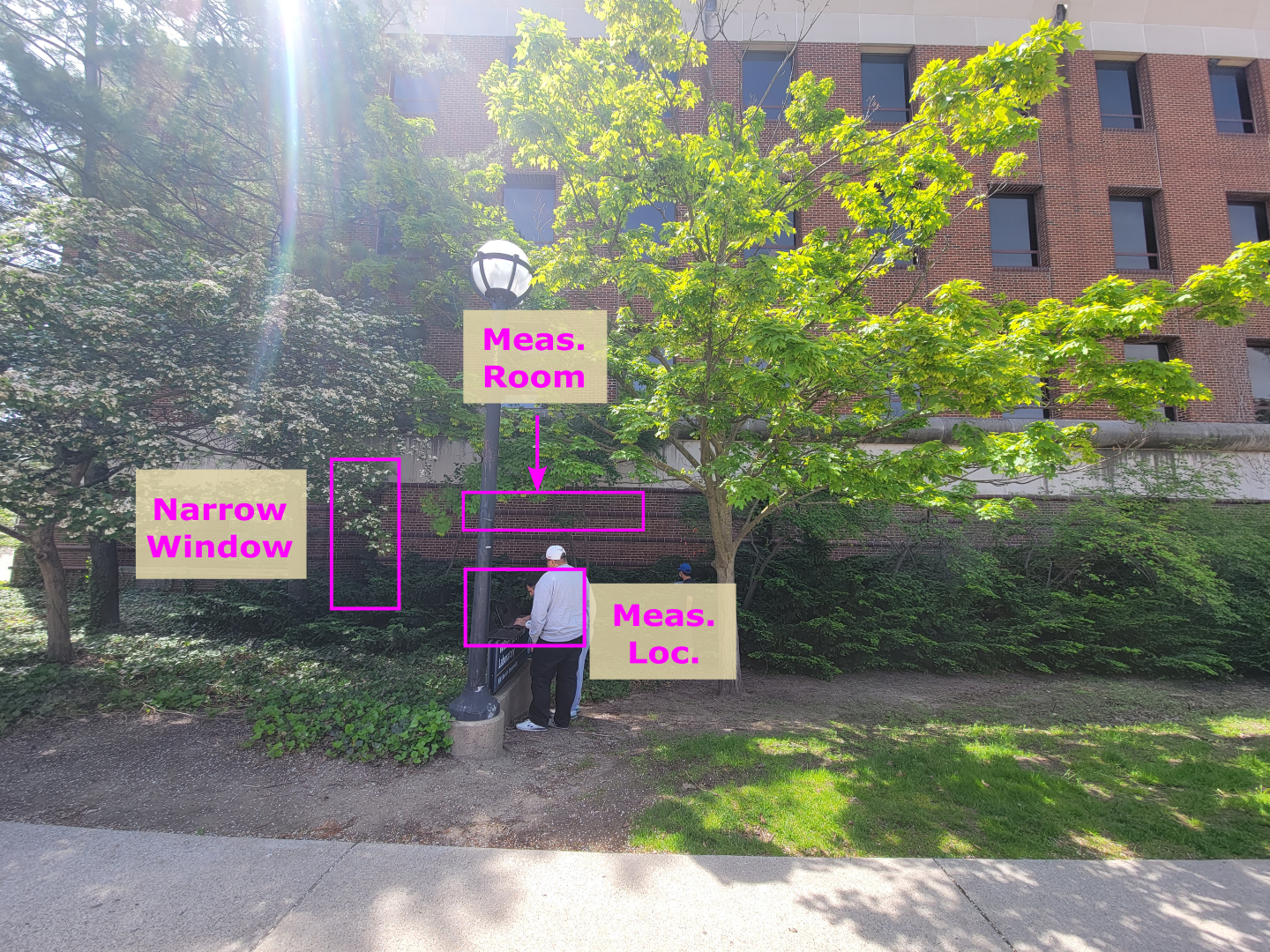}
        \label{Fig:chemistryout}}
     \end{subfigure} 
    \vspace{-1em}
    \caption{Measurement and AP locations for FL2.}
    \label{Fig:meas_locs_closed}
    \vspace{-1.5em}
\end{figure}

\begin{figure*}
     \centering
      \begin{subfigure}[Driving measurements.]{  \centering
    \includegraphics[scale= 0.42]{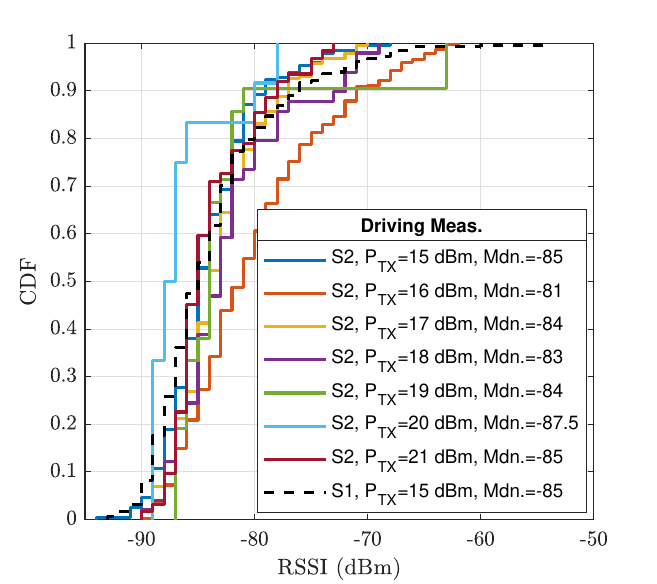}
    \label{Fig:cdf_rss_driving}}
    \end{subfigure} \hspace{-4mm}
     \begin{subfigure}[Walking measurements.]{
         \centering
         \includegraphics[scale= 0.42]{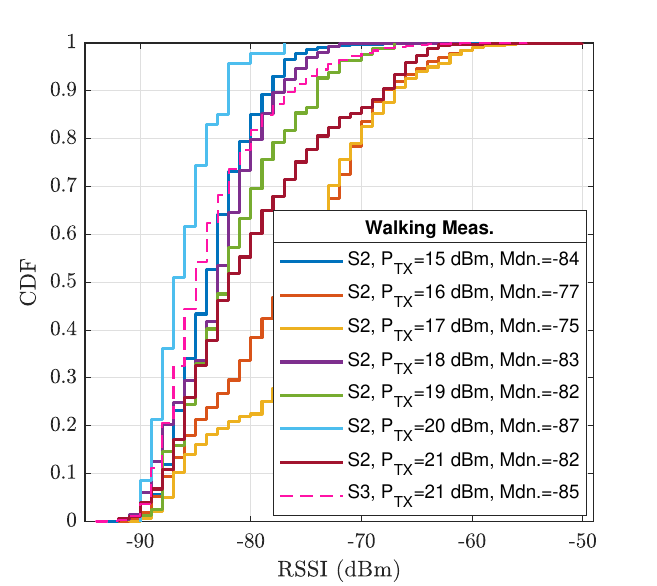}
        \label{Fig:cdf_rss_walking}}
     \end{subfigure}  \hspace{-4mm} 
      \begin{subfigure}[RSSI for each 80 MHz channel, S2 (driving).]{  \centering
    \includegraphics[scale= 0.42]{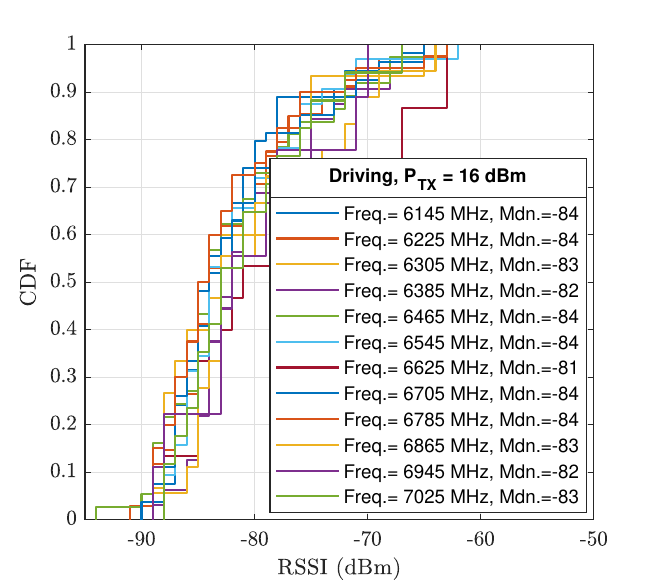}         \label{Fig:cdf_rss_driving_80}}
    \end{subfigure}  \vspace{-1em}
    \caption{CDF of outdoor RSSI, driving and walking. S1: MCA in Jan., S2: MCA in May, S3: RA in May.}
    \label{Fig:rssi_plots}
    \vspace{-1em}
\end{figure*}

\begin{figure}
     \centering
     \begin{subfigure}[Main campus area (MCA).]{  \centering
    \includegraphics[width=0.22\textwidth]{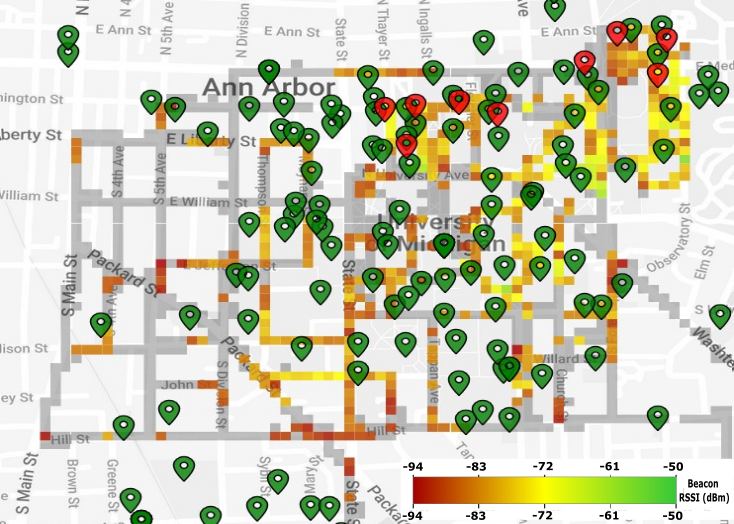}
         \label{Fig:heatmap_maincampus}}
    \end{subfigure} 
     \begin{subfigure}[Residential area (RA).]{
         \centering
\includegraphics[width=0.22\textwidth]{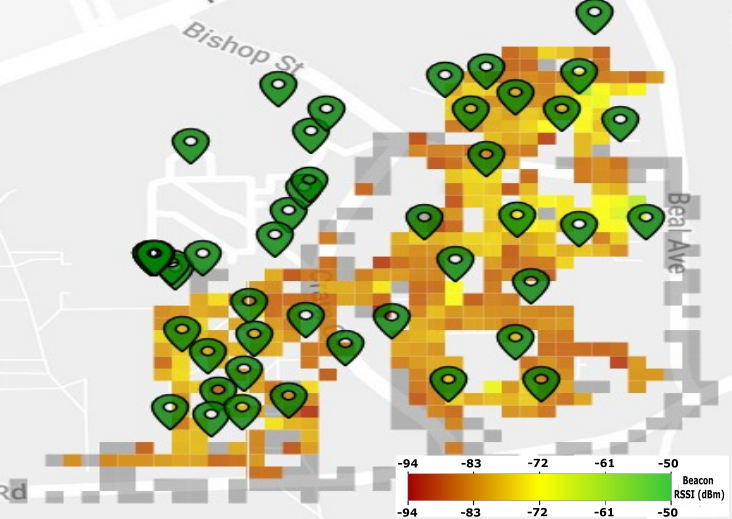}
        \label{Fig:heatmap_nw}}
     \end{subfigure}  \vspace{-1em}
    \caption{Outdoor RSSI heatmap for MCA and RA. Green pins: buildings with Wi-Fi 6E APs. Red pins: Drone experiment locations.}
    \label{Fig:heatmap_all}
    \vspace{-1em}
\end{figure}

\textit{Drone Measurements:} There are five active, fixed links in the MCA, as shown by the black lines in Fig. \ref{Fig:reg_campus2}. Three of these links have their transmitters (\textit{i.e.}, Tx1, Tx2 and Tx3 in the figure) located within the MCA, while the transmitters of the other two links (Tx4 and Tx5) are positioned at a significant distance away from the campus. Rx4 is the only receiver (Rx) on campus but the link direction is away from the buildings with dense deployments. Nine buildings, indicated by the orange pins in Fig. \ref{Fig:reg_campus2}, were chosen for drone measurements due to their proximity to Links 1 and 2, operating at center frequencies 7037.5 MHz and 6212.065 MHz with bandwidths of 25 MHz and 56 MHz respectively \cite{link1,link2}.
Table \ref{tab:dronebuildinginfo} provides information on the height of these buildings and the number of Wi-Fi 6E LPI APs deployed in each. On average, we assume two BSSIDs per AP in 6 GHz as determined by UMich ITS. \color{black} The drone measurements were conducted during daylight hours over a period of three days. As shown in Fig.~\ref{Fig:meas_locs_drone}, a Samsung S22+ smartphone with SigCap was tied to the drone for data collection. The drone moved vertically up and down, parallel to the wall of a given building. 

\textit{Driving Measurements:} In order to validate the driving measurements conducted in MC1, we replicated the same route as closely as possible. The measurements were carried out between 10:00 pm to 12:00 am, mirroring the timeframe of MC1 using the same 5 phones with SigCap.

\textit{Walking Measurements:}
The center of the campus, where Wi-Fi 6E is densely deployed, offers only pedestrian access. Hence RSSI measurements were collected in this area by walking with hand-held phones running SigCap (Fig.~\ref{Fig:reg_campus1}). 

\textit{Fixed Location 2 (FL2):} 
The measurement area is a conventional classroom on the first floor of a building, shown in Fig. \ref{Fig:meas_locs_closed}. The single AP in the room is center-mounted on the ceiling, and the room has a north facing exterior wall. The outdoor measurement location is 7 meters away from this wall due to trees obstructing closer access.

\section{Results \& Discussions}
We present statistical analyses of the measurements under different conditions. The discussions are categorized into two groups: (i) ground level driving \& walking measurements and (ii) aerial drone measurements. 

\subsection{Driving and Walking Measurements}

{\it Outdoor RSSI Levels:} Fig. \ref{Fig:reg_campus1} shows the map of outdoor beacon RSSI levels measured during the driving and walking campaigns. 
The minimum and maximum RSSI values measured across the MCA are -94 dBm and -62 dBm for the driving measurements, and -92 dBm and -55 dBm for the walking measurements, respectively. Transmit power levels ranging from $P_{TX}= 15$ dBm to $P_{TX}= 21$  dBm were observed within the MCA, with $P_{TX}= 16$ dBm being the most frequently used. $73\%$ and $95\%$ of the RSSI measurements were with $P_{TX} \leq 18$ dBm for the driving and walking measurements in the MCA, respectively. 

Statistical analyses of the measurements in the MCA and RA, using cumulative distribution function (CDF) plots of the measured RSSI at different transmit power levels, are shown in Fig. \ref{Fig:rssi_plots}.  Fig. \ref{Fig:cdf_rss_driving} shows the CDF of driving measurements within the MCA for MC1 (S1) and MC2 (S2). While $P_{TX}$ represents the transmit power for the AP, the maximum power of the 20 MHz beacon frames is $18$ dBm for $P_{TX}\geq 18$ dBm as shown in Table \ref{tab:regulation_6e}. MC1 measurements showed a transmit power of $15$ dBm: this changed when we returned in May for MC2. The median outdoor RSSI level is -85 dBm for both S1 and S2 under $P_{TX}= 15$ dBm, while the highest median RSSI value is -81 dBm  for S2 under $P_{TX}= 16$ dBm due to being the most frequently used. 

Fig. \ref{Fig:cdf_rss_walking} shows the CDF of outdoor RSSI levels recorded during walking measurements (only in MC2) in the MCA (S2) and the RA (S3). A single transmit power of $P_{TX}= 21$ dBm was observed in the RA deployment, which is less dense than the MCA and hence each AP can transmit at a higher power without interference. This is still 3 dB less than the maximum allowed power of 24 dBm for 80 MHz channels. Due to the proximity of the walking measurement locations to the buildings, an increase of 1-9 dBm is observed for the median RSSI values in the walking measurements compared to the driving measurements in the case of S2. Fig. \ref{Fig:cdf_rss_driving_80} shows the results obtained for each of the 80 MHz channels with Tx power of $P_{TX}= 16$ dBm: all the channels exhibit similar behavior. 

Finally, Fig. \ref{Fig:heatmap_all} shows the outdoor RSSI heatmap for the MCA and RA. As expected, the areas with a high concentration of APs have higher outdoor RSSI levels compared to areas with fewer APs. The gray areas show the regions where the 6 GHz beacon frames were not captured at all.

{\it Appropriate Enabling Signal Level for C2C mode:} In the proposed C2C mode, clients that can receive an enabling signal from any Wi-Fi 6E AP can directly communicate with each other, at  STA LPI power levels,  bypassing the need for data transmission through the AP. Device sharing is an example application that benefits from the C2C mode, reducing air time occupancy and latency. While the intended use of C2C is to improve indoor performance, care must be taken to set an appropriate level for the enabling signal so that client devices that are outdoors do not transmit to each other. The proposals submitted to the FCC recommended using -86 dBm/20 MHz and -82 dBm/20 MHz as enabling signal levels \cite{Intel}. Based on our walking results above, where the median outdoor RSSI level varies between -75 dBm and -85 dBm, even a level of -82 dBm could trigger > 50\% of outdoor devices to communicate with each other, which is not desirable. Hence, further measurements and analyses should be performed to determine an appropriate enabling signal level for C2C that minimizes the probability of interference.

\begin{figure}
     \centering
    \includegraphics[scale = 0.38]{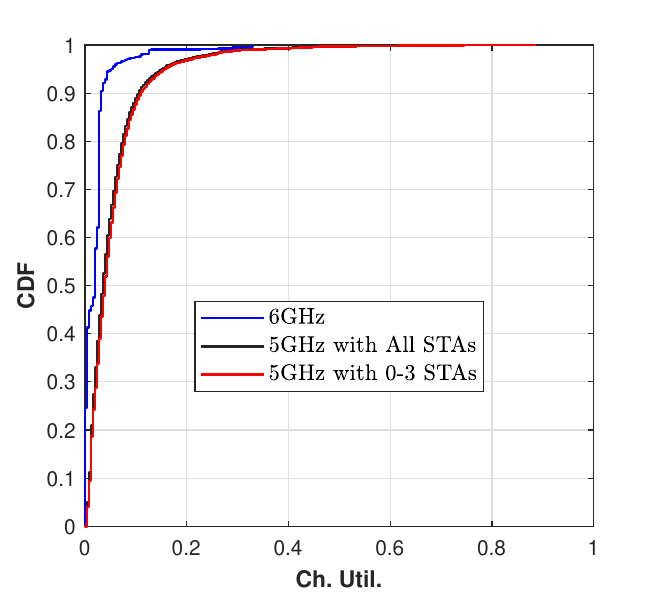} \vspace{-1em}
    \caption{Primary channel utilization for S2.}
    \label{Fig:meas_chutil}
     \vspace{-1.5em}
\end{figure}

\begin{figure}
     \centering
     \begin{subfigure}[Main campus area (MCA).]{  
     \centering
    \includegraphics[scale= 0.36]{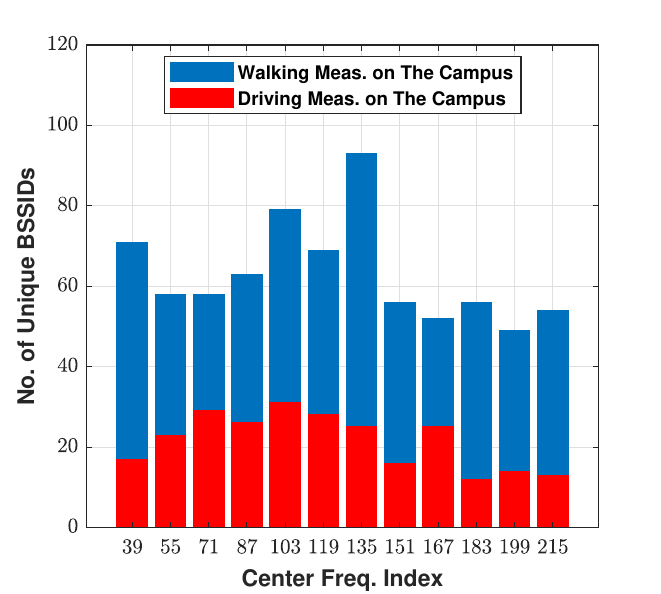}
         \label{Fig:bssidcampus}}
    \end{subfigure} \hspace{-7mm}
     \begin{subfigure}[Residential area (RA).]{
         \centering
         \includegraphics[scale= 0.36]{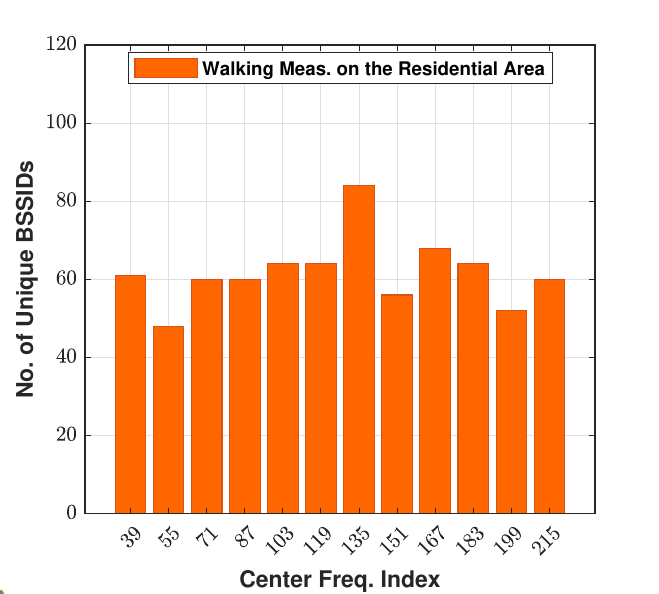}
        \label{Fig:bssidnw}}
     \end{subfigure} \vspace{-1em}
    \caption{Number of unique BSSIDs at MCA and RA. }
    \label{Fig:meas_locs1}
    \vspace{-1.5em}
\end{figure}

{\it Channel Utilization and Number of Unique BSSIDs:} Channel utilization and number of unique BSSIDs observed outdoors help to understand the potential interference impact of a dense deployment. A higher channel utilization and larger number of unique BSSIDs on a particular frequency point to increased potential for interference on that frequency. Fig.~\ref{Fig:meas_chutil} shows the CDFs of primary channel utilization at 5 GHz and 6 GHz for UMich SSIDs during walking measurements in the MCA. 6 GHz usage is still sparse, and during our measurements, a maximum of 3 devices were seen connected to a single BSSID at a
particular time instant. However, we observe from our 5 GHz measurements
that primary channel utilization is not very dependent on number of connected devices since the main
contributor to primary channel utilization is beacons and other management frames.

Fig. \ref{Fig:meas_locs1} shows the number of unique BSSIDs in each 80 MHz channel observed in the MCA and RA, demonstrating a similar pattern for the walking measurements in both areas and, as expected, a reduced number in the driving measurements in the MCA. The key takeaway from this result is that while there is a slightly higher number of unique BSSIDs observed outdoors on channel 135 in both areas, overall, all channels are used relatively uniformly, thus reducing the probability of interference to an outdoor fixed link that overlaps with a particular 80 MHz channel.

\begin{figure}
    \centering
    \includegraphics[scale = 0.40]{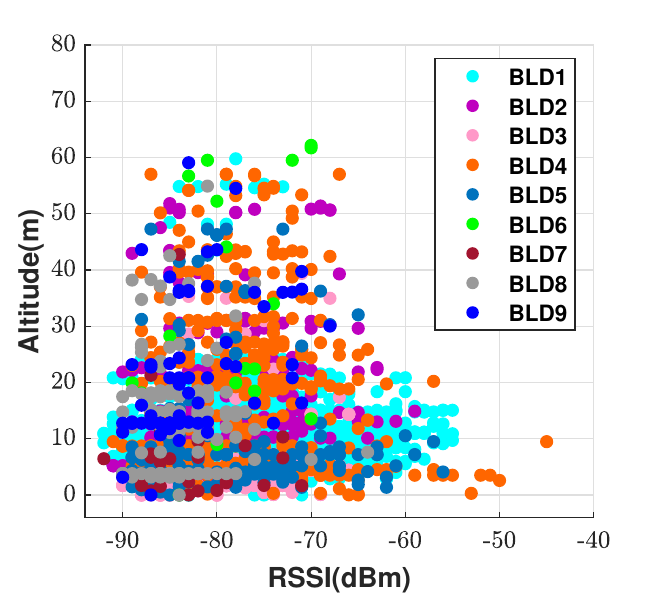} \vspace{-1em}
\caption{RSSI vs. altitude for drone measurements.}
    \label{Fig:drone_meas}
    \vspace{-2em}
\end{figure}

\begin{figure}
     \centering
     \begin{subfigure}[$N_{BSSID} 
 = 368 $]{  
     \centering
    \includegraphics[scale= 0.36]{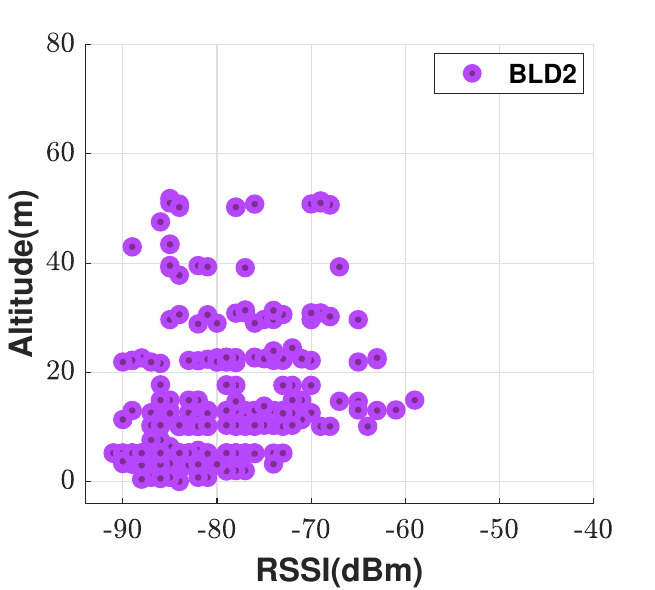}
         \label{Fig:mlb}}
    \end{subfigure} \hspace{-8mm}
     \begin{subfigure}[$N_{BSSID}  
 = 800 $]{
         \centering
         \includegraphics[scale= 0.36]{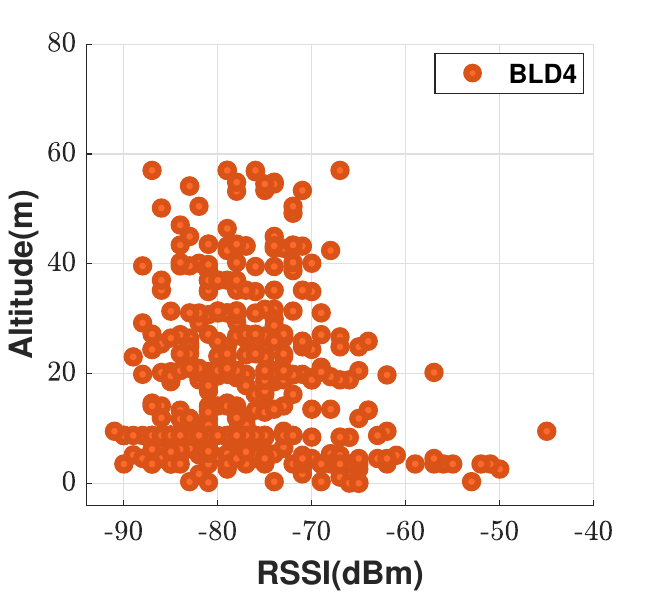}
        \label{Fig:ds}}
     \end{subfigure} \\ \vspace{-1em}
       \begin{subfigure}[$N_{BSSID} 
 = 68$]{  
     \centering
    \includegraphics[scale= 0.36]{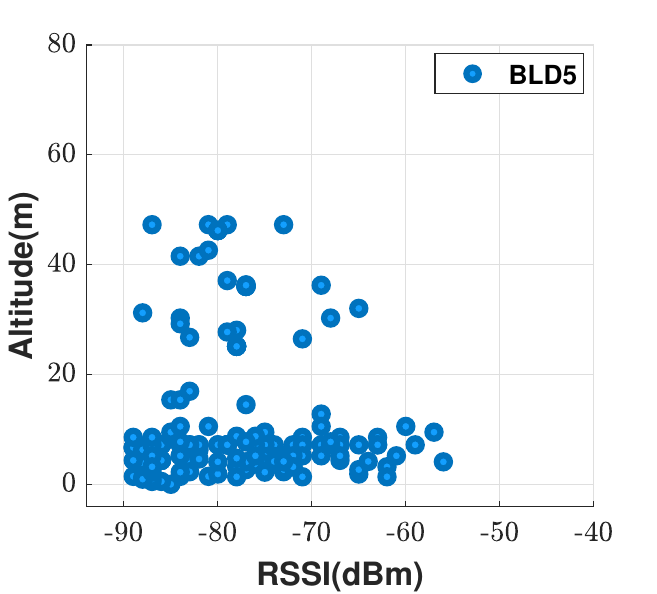}
         \label{Fig:hilla}}
    \end{subfigure}\hspace{-5mm}
     \begin{subfigure}[$N_{BSSID} 
 = 92 $]{
         \centering
         \includegraphics[scale= 0.36]{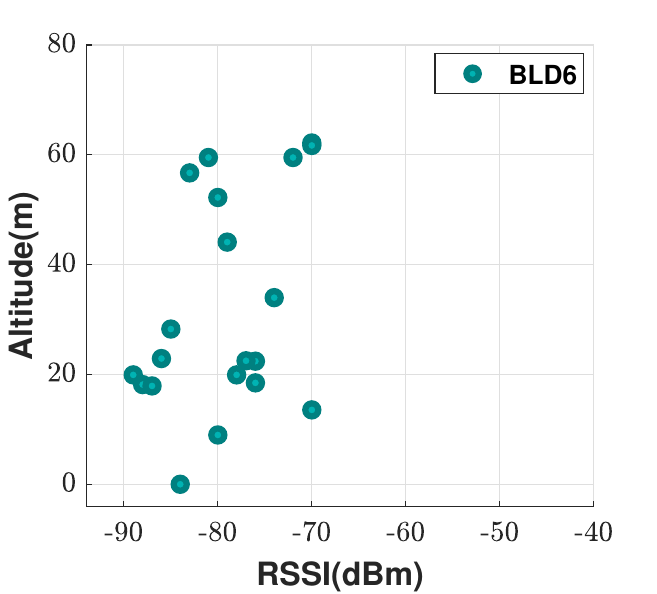}
        \label{Fig:pc}}
     \end{subfigure} \vspace{-1em}
    \caption{RSSI vs. altitude wrt. the number of BSSIDs.}
    \label{Fig:meas_locs_four}
    \vspace{-1.5em}
\end{figure}

\subsection{Drone Measurements}
Driving and walking measurements obtained at ground level alone do not offer a comprehensive understanding of the interference potential in the 6 GHz band since most outdoor fixed links are deployed at higher altitudes. Hence, the drone experiments provide insights into the RSSI levels as a function of altitude.

{\it Outdoor RSSI vs. Altitude:} Fig.~\ref{Fig:drone_meas} summarizes the RSSI measured at different altitudes near the nine buildings listed in Table \ref{tab:dronebuildinginfo}
The observed range of RSSI is between -93 dBm and -55 dBm. RSSI values greater than -60 dBm were not observed above a height of  20m.  Above a height of 30m, the RSSI values are less than -68 dBm. 

In order to provide an in-depth analysis of the relationship between RSSI and factors such as number of Wi-Fi 6E APs, construction material, and altitude, Fig. \ref{Fig:meas_locs_four} shows RSSI vs. altitude for four representative buildings: BLD2, BLD4, BLD5 and BLD6 with 368, 800, 68 and 92 BSSIDs respectively, as shown in Table \ref{tab:dronebuildinginfo}. BLD2 and BLD4 have many more APs compared to the other two. From Fig.~\ref{Fig:mlb} and Fig.~\ref{Fig:ds} we see that the drone measurements near BLD2 and BLD4 provide a larger number of data samples up to 60m compared to BLD5 and BLD6 which have fewer APs. However, there is an uniform decrease in the number of samples and RSSI with increase in altitude for all 4 buildings. Despite having fewer APs than BLD6, there are more data samples observed near BLD5 with higher RSSI: this is because unlike most buildings on campus, BLD5 is a historical building with single pane windows, resulting in lower loss.

\begin{figure}
     \centering
     \begin{subfigure}[-45 dBm RSSI outdoor.]{ 
     \centering
    \includegraphics[scale= 0.070]{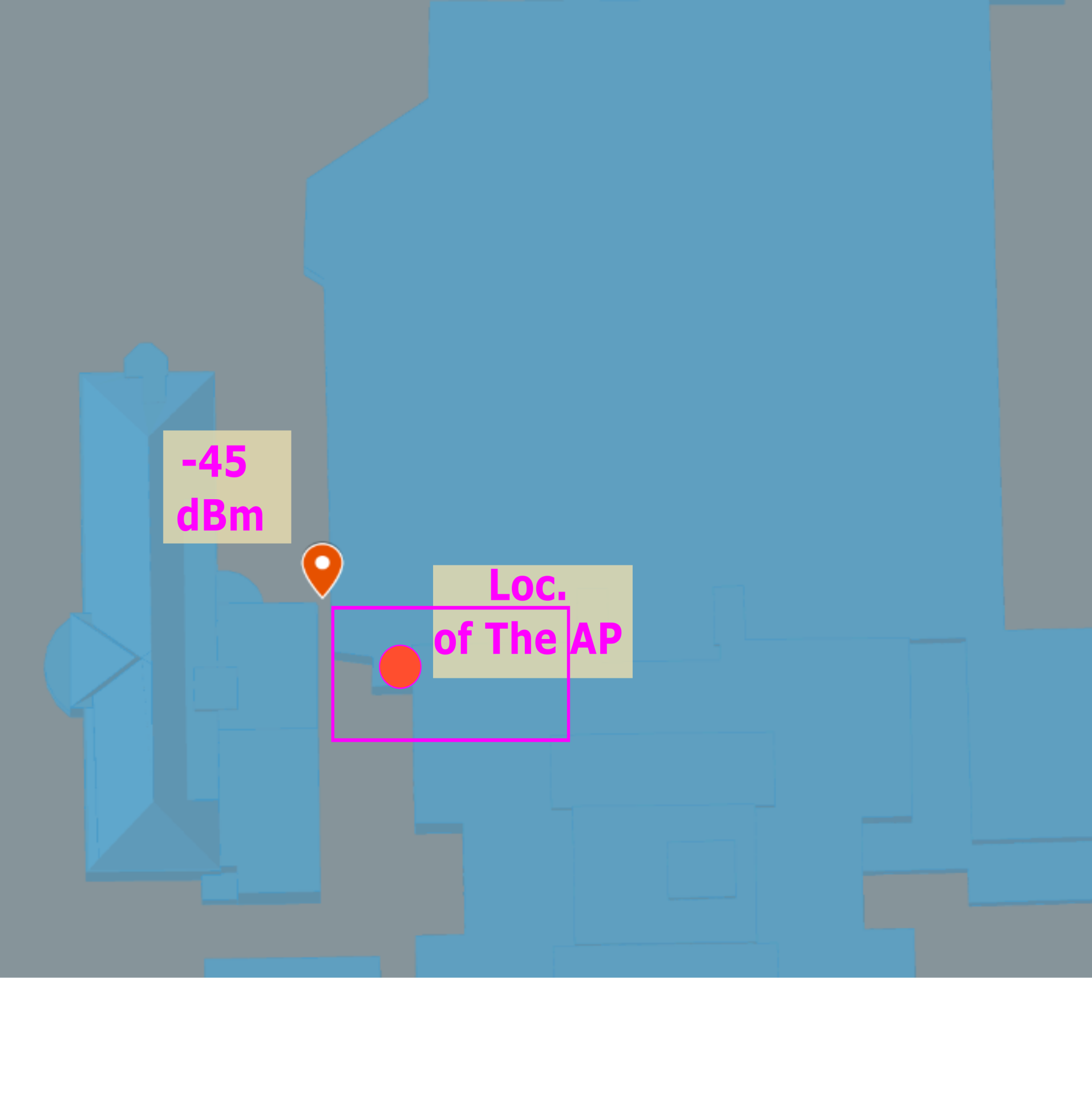}
         \label{Fig:APLOCDS}}
    \end{subfigure} \hspace{-7mm}
     \begin{subfigure}[All nine buildings.]{ 
     \centering
    \includegraphics[scale= 0.38]{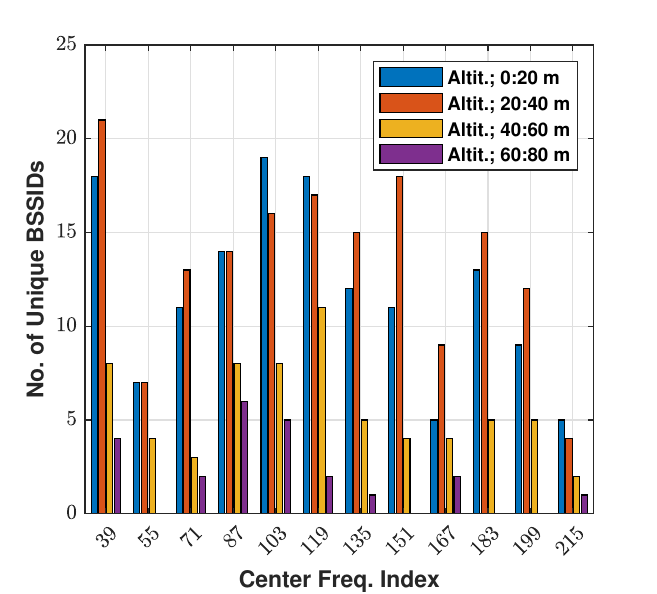}
         \label{Fig:DRONE_BSSID1}}
    \end{subfigure} \\ \vspace{-1em}
     \begin{subfigure}[BLD4.]{ 
     \centering
    \includegraphics[scale= 0.38]{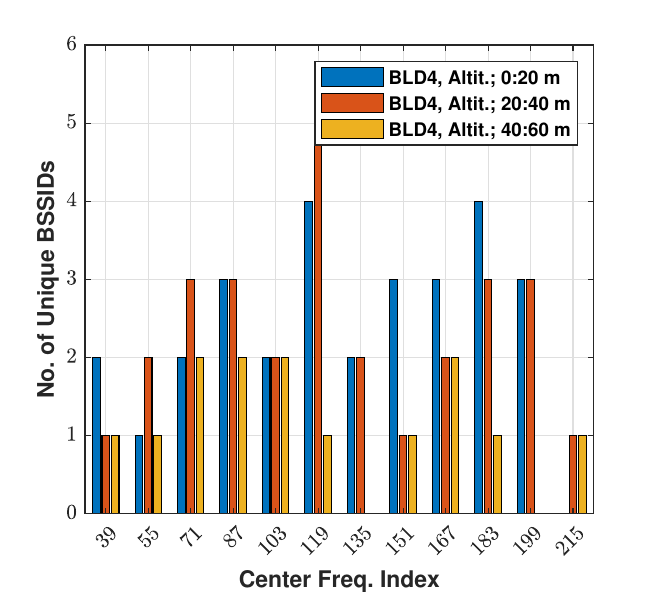}
         \label{Fig:DS_BSSID}2}
    \end{subfigure} \hspace{-8mm}
     \begin{subfigure}[CDF of RSSI.]{
         \centering
         \includegraphics[scale= 0.38]{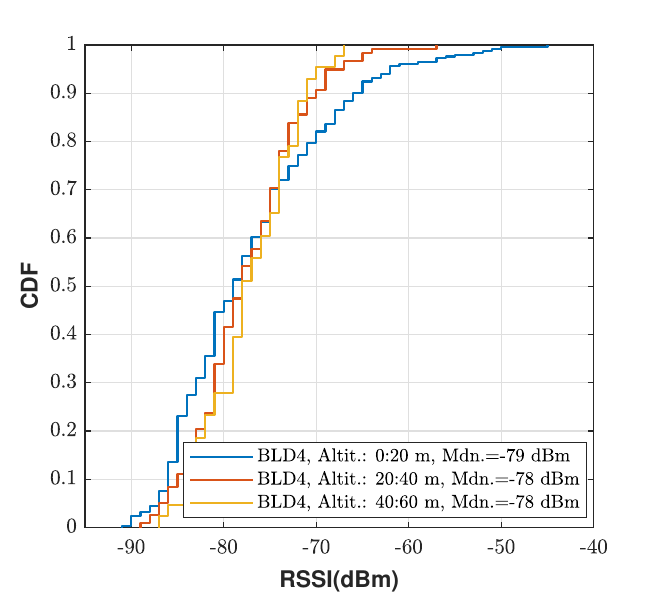}
        \label{Fig:DS_CDF}}
     \end{subfigure}
      \vspace{-1em}
    \caption{Number of unique BSSIDs vs. altitude.}
    \label{Fig:ds2}
    \vspace{-2em}
\end{figure}

{\it Number of Unique BSSIDs:} 
Fig.~\ref{Fig:ds} shows a high RSSI value of -45 dBm obtained at 10m near BLD4 which we investigate further. Fig.~\ref{Fig:APLOCDS} shows the relative location of this data sample and the corresponding BSSID/AP inside the building. The AP is in a room on the first floor and there is line-of-sight (LOS) through a corner window, resulting in the high outdoor RSSI measured at the outdoor location. It is important to note, however, that not all APs will contribute to significant signal emissions outdoors. In addition to the number of APs within a given building, the likelihood of these APs to LOS conditions through nearby windows plays a vital role in the resulting outdoor RSSI levels, and hence potential for interference. Figs. \ref{Fig:DRONE_BSSID1} and  \ref{Fig:DS_BSSID} illustrate the number of unique BSSIDs vs. altitude for the the nine buildings and for BLD4, respectively. Although the number of unique BSSIDs observed within the altitude range of $0-20$ m and $20-40$ m is fairly comparable, there is a noticeable decrease in the number of unique BSSIDs as the altitude range extends to $40-60$ m and $60-80$ m, thus indicating reduced potential for interference at higher altitudes. Finally, Fig. \ref{Fig:DS_CDF} shows the CDF of RSSI for BLD4. While the median outdoor RSSI values remain consistent across the three altitude intervals, there is a decrease in the maximum outdoor RSSI level as the altitude increases.

\begin{figure}
     \centering
     \begin{subfigure}[Link 1: RSSI on Channel 215.]{\centering
    \includegraphics[scale = 0.38]{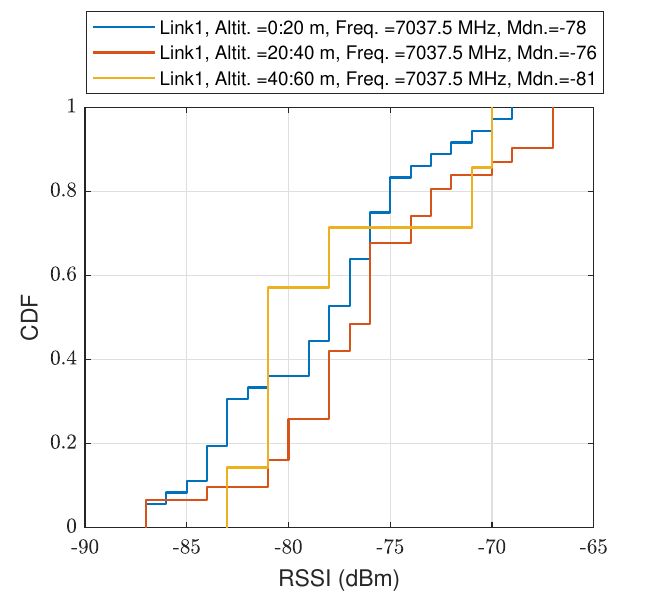}
\label{Fig:link1}}
    \end{subfigure} \hspace{-7mm}
     \begin{subfigure}[Link 2: RSSI on Channel 55.]{
         \centering
         \includegraphics[scale = 0.38]{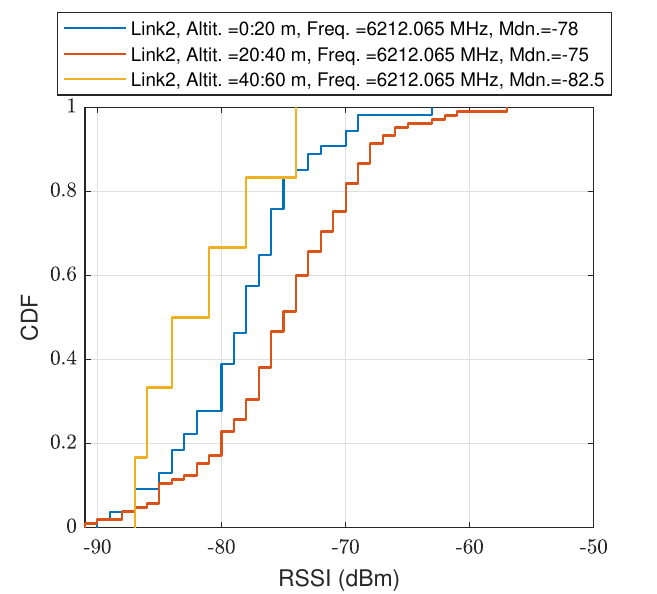}
\label{Fig:link2}}
     \end{subfigure}
      \vspace{-1em}
    \caption{CDF of drone RSSI measurements on channels overlapping Links 1 and 2.}
    \label{Fig:6ghz_fixed_link_perf}
     \vspace{-2em}
\end{figure}

\begin{figure}
     \centering
     \begin{subfigure}[Double-pane low-E window.]{\centering
    \includegraphics[scale = 0.38]{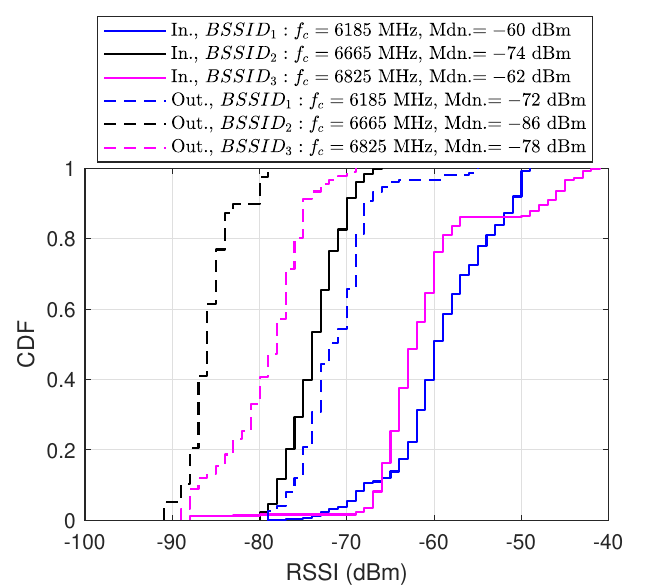}
\label{Fig:in_out_entry_loss_open}}
    \end{subfigure} \hspace{-7mm}
     \begin{subfigure}[Solid brick wall.]{
         \centering
         \includegraphics[scale = 0.38]{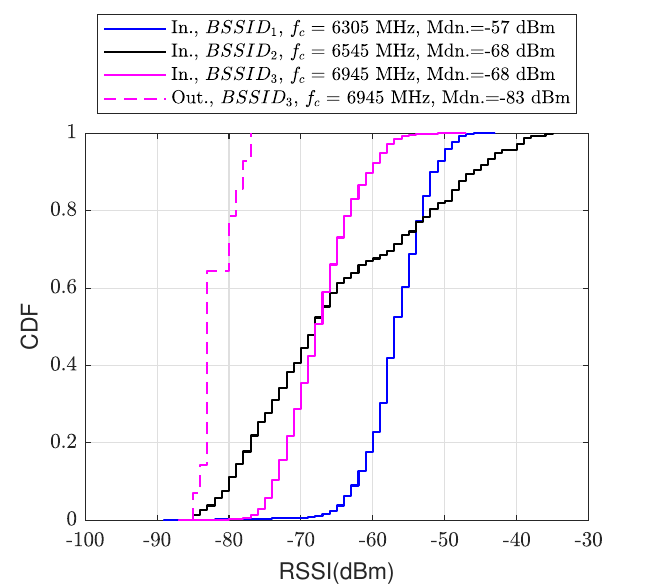}
\label{Fig:in_out_entry_loss_closed}}
     \end{subfigure} \vspace{-1em}
    \caption{BEL for fixed locations FL1  and FL2.}
    \label{Fig:in_out_entry_loss}
    \vspace{-2em}
\end{figure}

{\it Interference with fixed links:}
We evaluate the interference potential to Links 1 and 2 which overlap with Wi-Fi channels 215 and 55 respectively (Link 2 has < 1 MHz overlap with the edge of channel 39 which we ignore since the Wi-Fi signal drops off at the band-edge). Fig. \ref{Fig:6ghz_fixed_link_perf} shows the CDF of the RSSI on these channels at different altitudes. As the altitude increases, RSSI level decreases, thus reducing the interference potential to these links. To further evaluate the interference level, we calculate approximately the ratio of interference to noise power (I/N)  for these links as $ I/N = 10\log_{10}(BW_i/20)+ RSSI_{Outdoor}+G_{rx}-NF-PL$, where $BW_i$ is the link bandwidth, $G_{rx}$ is the Rx antenna gain, $NF$ is the noise floor and $PL$ is the free space path loss. These are computed from the link parameters in \cite{link1,link2}. We assume worst case conditions: highest outdoor RSSI measured of -68 dBm and
 -58 dBm for Links 1 and 2 respectively, in the main Rx beam. $I/N$ is calculated to be -72 dB for Link 1 and -66 dB for Link 2, much lower than the harmful interference threshold of $I/N = -6$ dB.
 Although Rx4 is located in the MCA, the link points away from the densely deployed region and thus we did not calculate the interference level at Rx4.

\subsection{Indoor-Outdoor BEL Measurements}

{\it BEL near a double-pane low-E Window:} Fig.\ref{Fig:in_out_entry_loss_open} shows the CDF of indoor and outdoor RSSI values for the fixed location FL1 which is the open area shown in Fig. \ref{Fig:meas_locs_jan}. We only consider RSSI measurements where the client devices are connected to the BSSID associated with the AP in the room,  which is one of the few APs with three BSSIDs. 
A 12 dB BEL is observed for $\text{BSSID}_{1}$ and $\text{BSSID}_{2}$, while $\text{BSSID}_3$ exhibits a higher entry loss of 16 dB.   

{\it BEL near a solid brick wall:} Fig. \ref{Fig:in_out_entry_loss_closed} shows the results obtained for the FL2 shown in Fig.~\ref{Fig:meas_locs_closed}. Inside the  measurements room, the devices were able to connect to $BSSID_1$ and $BSSID_2$.
However, these two BSSIDs were not detected outside due to the solid brick wall. $BSSID_3$ was observed outside since it is associated with the AP located in the adjacent room, which has a window pointing out towards the outdoor measurement location. Moreover, 391 APs, corresponding to 782 BSSIDs, are deployed in the entire building of which only 159 BSSIDs are observed within the measurement room, and only 8 of these i.e., $5\%$, are observed outside in this location, indicating a very high loss through the brick wall.

\section{Conclusions \& future research}

We conducted an extensive measurement campaign via drone, driving, walking, and indoor-outdoor measurements at the world's largest indoor Wi-Fi 6E deployment on the UMich campus, investigating the interference potential of densely deployed LPI APs. To the best of the authors' knowledge, this is the first such measurement campaign conducted on a real-world Wi-Fi 6E deployment. In-depth analyses of the relationship between outdoor RSSI levels and factors such as the number of APs, the positioning of the APs in relation to nearby windows, and altitude is provided.  Most LPI APs within a building cannot be received outdoors, but a few APs with LOS through windows can result in high outdoor RSSI levels in a very small number of locations, e.g. only $5\%$ of the indoor BSSIDs in one building are observed outdoors in a location near a solid brick wall.
The BEL near double-pane low-E windows was 12 - 16 dBm. Drone measurements show the number of unique BSSIDs and outdoor RSSI levels decrease with increasing altitude, further reducing interference potential. Based on median outdoor RSSI levels, further measurements and analyses are required to determine an appropriate enabling signal level for C2C mode. Future research will investigate collaborations with fixed-link providers to quantify interference at the incumbent receiver.

\section*{\centering{Acknowledgements}}
We thank the Information and Technology Services of UMich for their help throughout the measurement campaign and support of this research. 
The research was funded in part by NSF Grant\# CNS-2229387.

\bibliographystyle{ieeetr}
\bibliography{Paper}

\end{document}